\documentclass[aps,prd,nofootinbib]{revtex4}
\usepackage{amsmath,amssymb}
\usepackage[hyperfootnotes=false]{hyperref}

\begin{document}

\title{Dark photon portal into mirror world}

\author{Abdaljalel Alizzi}
\email{abdaljalel90@gmail.com}
\affiliation{Novosibirsk State University, Novosibirsk 630 090, Russia}

\author{Z.~K.~Silagadze}
\email{Z.K.Silagadze@inp.nsk.su}
\affiliation{Budker Institute of Nuclear Physics and Novosibirsk State 
University, Novosibirsk 630 090, Russia}

\begin{abstract}
Dark photons and mirror matter are well-motivated dark matter candidates. 
It is possible that both of them arose during the compactification and 
symmetry breaking scenario of the heterotic $E_8\times E_8$ string theory 
and are related to each other. In this case, dark photons can become a natural 
portal into the mirror world. Unfortunately, the expected magnitude of the 
induced interactions of ordinary matter with mirror matter is too small 
to be of phenomenological interest. 
\end{abstract}

\maketitle

\section{introduction}
There is overwhelming evidence for the existence of dark matter at all 
astrophysical length scales, from galactic to cosmological \cite{1}. At 
the same time, the true nature and composition of dark matter remain unknown. 
All evidence for the existence of dark matter so far is based on its 
gravitational effects. Many proposed models of dark matter assume other 
non-gravitational very weak interactions between dark matter particles and 
ordinary matter. This motivates dark matter direct-detection experiments 
\cite{2,3}. The sensitivity of such experiments has improved tremendously. 
However, no clear experimental evidence for the existence of dark matter 
particles has been provided yet from these experiments. 

A generic feature of string theory is the prediction of extra $U(1)$ gauge 
factors beyond the Standard Model group \cite{4}. This fact makes dark
photons well-motivated candidates  for dark matter \cite{5}. These dark
photons can kinetically mix with the Standard Model photon \cite{6} and thus
lead to potentially observable effects in astrophysical and cosmological
phenomena, as well as in laboratory experiments \cite{5,7}. 

Perhaps some comments about dark photons as dark matter candidate are 
appropriate here. Our usual intuition is that light vector bosons mediate 
forces, but cannot make up matter in our universe. However very light
dark photons can challenge this intuition. 

An interesting alternative to cold dark matter is the so-called fuzzy dark 
matter, consisting of very light bosons with extremely long de Broglie 
wavelengths \cite{7A,7B}. Inside galaxies, such dark matter can be modeled as 
a classical field and can lead to a variety of dark condensed matter physics 
phenomena \cite{7C}.

If relic abundance of dark photon dark matter is produced by quantum 
fluctuations during inflation, then its energy density $\Omega_B$ relative to 
the critical density today is estimated to be \cite{7D}
\begin{equation}
\Omega_B=\Omega_{DM}\sqrt{\frac{m_B}{6\times 10^{-6}~\mathrm{eV}}}\left (
\frac{H_I}{10^{14}~\mathrm{GeV}}\right )^2,
\label{A1}
\end{equation}
where the inflationary Hubble scale $H_I\le 10^{14}~\mathrm{GeV}$ is bounded 
by the absence of observation of primordial gravitational waves \cite{7C}. 
Therefore, in this scenario, dark photons from inflationary fluctuations can 
constitute the dominant part of the dark matter only if $m_B\ge \mu\mathrm{eV}$ 
\cite{7C}. However, other scenarios, when the dark photon dark matter
abundance is generated by instabilities in  a misaligned axion, operate over
a wider range of dark photon masses down to $10^{-20}~\mathrm{eV}$ \cite{7C}.

Another well-motivated candidate for dark matter is mirror matter. The mirror
partners of ordinary particles were first introduced by Lee and Yang
in their famous paper \cite{8} in an attempt to preserve the left-right 
symmetry of the world. Kobzarev, Okun and Pomeranchuk realized \cite{9} 
that mirror particles could make up a hidden sector of the world, which 
communicates with the visible world mainly through gravity. The idea was 
rediscovered in the modern context of renormalizable gauge theories by Foot, 
Lew and Volkas \cite{10} and  revived in the context of neutrino oscillations 
in both mirror-symmetric \cite{11} and mirror-asymmetric \cite{12} forms. 
For recent reviews of the mirror matter theory and related references, see, 
for example, \cite{13,14,15,16}.

Mirror matter, like dark photons, can also be motivated (with some 
caveats) by string theory. One of the heterotic string theories, which makes 
it possible to construct a coherent quantum theory that unifies all 
interactions, including gravity, is based on the group $E_8\times E_8$ 
\cite{17,18}. The second $E_8$ after compactification can lead to the 
existence of shadow matter, which interacts only gravitationally with ordinary 
matter \cite{19}. Mirror matter is a special case of shadow matter when the 
symmetry breaking patterns of the second $E_8$ exactly mirror the patterns of 
the first $E_8$, which leads to an exciting possibility of a mirror world with 
invisible stars, planets and galaxies \cite{9,20}. 

Some of the caveats mentioned above are as follows. In the $E_8\times E_8$
heterotic string model we have an exact symmetry between two $E_8$-s at the 
stringy scale when the basic constituents of the universe are strings living
in ten-dimensional space-time (as dictated by anomaly cancellation). Then 
universe cools and the extra six dimensions compactify in the spirit of 
Kaluza-Klein theory, so that the space-time becomes $M_4\times K_6$, where
$M_4$ is a familiar 4-dimensional space-time and $K_6$ is some compact 
6-dimensional manifold. Usually one demands  an unbroken $N=1$ supersymmetry 
in four dimensions and in this case the natural choice for $K_6$ is 
a Calabi-Yau manifold \cite{20AA}.

We emphasize that nothing in the theory itself requires such compactification. 
This is just a phenomenological assumption, very popular around the turn of 
the millennium (but perhaps less popular today due to the lack of experimental 
evidence for supersymmetry).  

Even if one assumes this particular type of compactification, the number of
possible phenomenologically viable compactifications are huge. For complete 
intersection Calabi-Yau threefolds the number of string theory standard models
is estimated to be $10^{23}$ (a mole of standard models in Tristan H\"{u}bsch's
apt characterization), and for the class of Calabi-Yau hypersurfaces in toric 
varieties, the estimated number of standard models is truly terrifying 
$10^{723}$ \cite{20BB}. It is safe to say that it is currently unclear which 
path leads from string theory to low-energy physics of the Standard Model, 
and it is also unclear whether string theory actually describes nature.

The standard Calabi-Yau compactification breaks the mirror symmetry between
two  $E_8$-s because the $SU(3)$ holonomy group of the Calabi-Yau manifold is
chosen in one of the $E_8$-factors breaking it to $E_6$, while the second
$E_8$ remains unbroken.  

One might think that an alternative possibility is to embed the SU(3) holonomy 
in a diagonal  subgroup of $E_8\times E_8$ and in this way maintain the 
symmetry between the visible and shadow matters. Unfortunately, such Calabi-Yau 
spaces have not been constructed \cite{20CC}.

Therefore, in what follows, for definiteness, we will focus on mirror matter, 
but the scenario considered is actually applicable for a wider class of shadow 
matter theories, which do not necessarily assume exact mirror symmetry between 
ordinary and shadow particles. 
  
Since both dark photons and mirror matter may have their origins in string 
theory, it can be assumed that they are related in some way\footnote{In 
addition to string theory, both mirror matter and  dark photons were motivated
in a five-dimensional unification of gravitation and electromagnetism using 
a degenerate metric \cite{20A}.}. In this short note, we present the simplest 
model of this type: the visible and shadow sectors, connected only (except 
gravity) by dark photon messengers.

Although our initial motivation was a hint from string theory, we do not 
really need this hypothesis for what follows. Equally successfully can one 
motivate such a picture from the $E_6\times E_6$ grand unification theory with 
symmetric or asymmetric symmetry breaking patterns of two $E_6$-s \cite{20AB}.

\section{Mirror world with dark photons}
Calabi-Yau compactifications of the heterotic $E_8\times E_8$ superstring 
model naturally leads to the gauge group $E_6$ in the visible sector
\cite{20,21}. Further symmetry breaking patterns such as 
$E_6\to SO(10)\times  U(1)$ or $E_6\to SU(5)\times SU(2)\times U(1)$ can 
introduce the gauge group $U(1)$ of dark photons into play\footnote{
$E_8\times E_8$ superstring phenomenology, including shadow world was 
discussed, for example,  in \cite{21A}.}.

It is usually assumed that after Calabi-Yau compactification, the gauge
group in the hidden sector is different from the visible sector gauge group, 
since, as noted in~\cite{19}, a fully symmetric hidden sector contradicts 
observations, in particular Big Bang nucleosynthesis constraints. However,  
the limitations from the Big Bang nucleosynthesis can be avoided if the mirror 
sector has a lower temperature than the ordinary one, and  an inflationary 
scenario can be envisaged that could explain the different initial 
temperatures of the two sectors \cite{16}. As a result, although the 
microphysics of these two sectors are identical, their macroscopic properties 
in relation to the most important epochs, such as baryogenesis, 
nucleosynthesis, etc., will be completely different \cite{16}.

Therefore, we assume that it is possible to maintain either an exact mirror 
symmetry or its broken version between the visible and hidden sectors, 
starting from the  scale of the $E_6\times E_6$ grand unification and up to 
the electroweak scale. Then during, say, $E_6\times E_6\to(SO(10)\times  U(1))
\times(SO(10)\times  U(1))$ symmetry breaking stage the two $U(1)$-s of the 
dark photon and mirror dark photon can get mixed, for example, through the 
Higgs portal \cite{22}.

These considerations prompt us to consider a scenario when mirror matter (or 
its shadow analogue) predominates in dark matter, and dark photons play 
merely the role of a portal between the visible and hidden sectors. We 
describe this situation by an effective Lagrangian density 
\begin{equation}
{\cal{L}}=L+\tilde L-\frac{\epsilon_2}{2}F_{b\mu\nu}\tilde F^{b\mu\nu}-
e_aJ_\mu A^{a\mu}-e_a\tilde J_\mu \tilde A^{a\mu},
\label{eq1}
\end{equation}
where tilde denotes mirror fields, $J_\mu$ and $\tilde J_\mu$ are
ordinary and mirror electric currents, and
\begin{equation}
\begin{aligned}
& L=-\frac{1}{4}F_{a\mu\nu}F^{a\mu\nu}-\frac{1}{4}F_{b\mu\nu} 
F^{b\mu\nu}+\frac{1}{2}m_b^2A_{b\mu} A^{b\mu}-\frac{\epsilon_1}{2}\,
F_{a\mu\nu}F^{b\mu\nu}, \\
&  \tilde L=-\frac{1}{4}\tilde F_{a\mu\nu}\tilde F^{a\mu\nu}-
\frac{1}{4}\tilde F_{b\mu\nu} \tilde F^{b\mu\nu}+\frac{1}{2}m_b^2\tilde A_{b\mu} 
\tilde A^{b\mu}-\frac{\epsilon_1}{2}\,\tilde F_{a\mu\nu}\tilde F^{b\mu\nu}.
\end{aligned}
\label{eq2}
\end{equation}
The dark photon masses can arise (in each sector), for example, by 
St\"{u}ckelberg mechanism \cite{23}, or by the usual Higgs mechanism.

The Lagrangian (\ref{eq1}) can be diagonalized in three steps. First, we
introduce the fields $A_\mu$ and $B^\prime_\mu$ along their mirror counterparts
via
\begin{equation}
\begin{aligned}
& A_{a\mu}=A_\mu-\frac{\epsilon_1}{\sqrt{1-\epsilon_1^2}}\,B^\prime_\mu,\;\;\;
A_{b\mu}=\frac{1}{\sqrt{1-\epsilon_1^2}}\,B^\prime_\mu, \\
& \tilde A_{a\mu}=\tilde A_\mu-\frac{\epsilon_1}{\sqrt{1-\epsilon_1^2}}\,
\tilde B^\prime_\mu,\;\;\;\tilde A_{b\mu}=\frac{1}{\sqrt{1-\epsilon_1^2}}\,
\tilde B^\prime_\mu.
\label{eq3}
\end{aligned}
\end{equation}
In terms of these fields, Lagrangians (\ref{eq2}) take the diagonal forms
\begin{eqnarray} &&
L=-\frac{1}{4}F_{\mu\nu}F^{\mu\nu}-\frac{1}{4}B^\prime_{\mu\nu}B^{\prime\mu\nu}+
\frac{1}{2}\,\frac{m_b^2}{1-\epsilon_1^2}\,B^\prime_{\mu} B^{\prime\mu},
\nonumber \\ &&
\tilde L=-\frac{1}{4}\tilde F_{\mu\nu}\tilde F^{\mu\nu}-\frac{1}{4}
\tilde B^\prime_{\mu\nu}\tilde B^{\prime\mu\nu}+\frac{1}{2}\,
\frac{m_b^2}{1-\epsilon_1^2}\,\tilde B^\prime_{\mu} \tilde B^{\prime\mu}.
\label{eq4}
\end{eqnarray}
However, visible and hidden gauge fields still remain kinematically 
intermixed thanks to the $-\frac{\epsilon_2}{2(1-\epsilon^2_1)}
B^\prime_{\mu\nu}\tilde B^{\prime\mu\nu}$ term. This term can be diagonalized by 
the transformation
\begin{equation}
B^\prime_{\mu}=\frac{1}{\sqrt{2}}\left (B^{\prime\prime}_{\mu}+\tilde 
B^{\prime\prime}_{\mu}\right ),\;\;\;\tilde B^\prime_{\mu}=\frac{1}{\sqrt{2}}\left 
(B^{\prime\prime}_{\mu}-\tilde B^{\prime\prime}_{\mu}\right ).
\label{eq5}
\end{equation}
Being an orthogonal transformation, (\ref{eq5}) does not spoil the
diagonality of the sum of Lagrangians (\ref{eq4}). However, we end up
with incorrect coefficients of the kinetic $B^{\prime\prime}_{\mu\nu} 
B^{\prime\prime\mu\nu}$ and $\tilde B^{\prime\prime}_{\mu\nu} \tilde B^{\prime\prime\mu\nu}$ 
terms. To restore the correct normalization of these terms, we rescale the 
fields:
\begin{equation}
B^{\prime\prime}_{\mu}=\sqrt{\frac{1-\epsilon_1^2}{1-\epsilon_1^2+\epsilon_2}}\,
B_\mu,\;\;\;\tilde B^{\prime\prime}_{\mu}=\sqrt{\frac{1-\epsilon_1^2}{1-\epsilon_1^2
-\epsilon_2}}\,\tilde B_\mu .
\label{eq6}
\end{equation}
Finally, in terms of mass-eigenstate physical fields (ordinary photon $A_\mu$,
mirror photon $\tilde A_\mu$, dark photon $B_\mu$ and mirror dark photon 
$\tilde B_\mu$) our original Lagrangian density (\ref{eq1}) takes the form
\begin{equation}
{\cal{L}}=-\frac{1}{4}F_{\mu\nu}F^{\mu\nu}-\frac{1}{4}B_{\mu\nu}B^{\mu\nu}
-\frac{1}{4}\tilde F_{\mu\nu}\tilde F^{\mu\nu}-\frac{1}{4}\tilde B_{\mu\nu}
\tilde B^{\mu\nu}+ \frac{1}{2}\,\mu^2B_\mu B^\mu+\frac{1}{2}\,\tilde \mu^2
\tilde B_\mu \tilde B^\mu+{\cal{L}}_{\mathrm{int}}.
\label{eq7}
\end{equation}
Here, as usual,  $F_{\mu\nu}=\partial_\mu A_\nu-\partial_\nu A_\mu$, $B_{\mu\nu}=
\partial_\mu B_\nu-\partial_\nu B_\mu$, and 
\begin{equation}
\mu^2=\frac{m_b^2}{1-\epsilon_1^2+\epsilon_2},\;\;\;
\tilde\mu^2=\frac{m_b^2}{1-\epsilon_1^2-\epsilon_2}.
\label{eq8}
\end{equation}
The most interesting is the interaction term
\begin{equation}
{\cal{L}}_{\mathrm{int}}=-e_aJ_\mu\left [A^\mu-\frac{\epsilon_1}{\sqrt{2}}\left (
\frac{B^\mu}{\sqrt{1-\epsilon_1^2+\epsilon_2}}+
\frac{\tilde B^\mu}{\sqrt{1-\epsilon_1^2-\epsilon_2}}\right )\right ]-
e_a\tilde J_\mu\left [\tilde A^\mu-\frac{\epsilon_1}{\sqrt{2}}\left (
\frac{B^\mu}{\sqrt{1-\epsilon_1^2+\epsilon_2}}-
\frac{\tilde B^\mu}{\sqrt{1-\epsilon_1^2-\epsilon_2}}\right )\right ].
\label{eq9}
\end{equation}
As we can see, the effect of $\sim \epsilon_2$ mixing term in (\ref{eq1}) is
twofold: it removes the degeneracy between dark photon and mirror dark photon,
resulting in mass splitting
\begin{equation}
\mu^2-\tilde\mu^2=-\frac{2\epsilon_2\,m_b^2}{(1-\epsilon_1^2)^2-\epsilon_2^2}
\approx -2\,\epsilon_2\,m_b^2,
\label{eq10}
\end{equation}
and interconnects the visible and hidden sectors (if $\epsilon_2=0$, the 
interconnection implied by (\ref{eq9}) is unphysical and can be rotated away 
by an orthogonal transformation of the physical fields).

\section{Concluding remarks}
Previously, several different ``portals'' were considered that connect the 
ordinary and mirror worlds: photon portal (photon-mirror photon 
oscillations) \cite{24,25}, neutrino portal \cite{11,12,26A,26}, Higgs portal
\cite{27,27A,27B}, axion portal \cite{28,28A,28B}, neutron portal 
(neutron-mirror neutron oscillations) \cite{29}\footnote{It was stated in 
\cite{29A} that the internal heating of a neutron star severely constraints 
the neutron portal. However, the internal heating of a neutron star was 
discussed earlier in \cite{29B}, and it was concluded that the constraints are 
not at all so severe: they are comparable to the direct experimental bounds
\cite{29C,29D} and thus still allow neutron oscillations much faster than the 
neutron decay.}. Even in the absence of all these portals, the visible and 
mirror sectors become interconnected (albeit very weakly) by quantum gravity 
effects \cite{30}.

The dark photon portal proposed in this note is, in our opinion, very 
natural. Combining two dark entities is also helpful in explaining
the dark matter mystery. For light dark photons to be the dominant part of 
dark matter, it is necessary to identify a well-motivated mechanism for their 
production. Although some production mechanisms have been proposed 
(misalignment mechanism, production due to fluctuations of the metric during 
inflation of the early universe, temperature-dependent instabilities in 
the hidden-sector (pseudo)scalar field coupled to a dark photon), they are 
not without difficulties \cite{31}.

In the case of mirror matter, there are no such difficulties, and mirror 
baryons could easily provide the dominant component of dark matter in 
the universe \cite{16}. For example, interactions leading to mixing of 
neutrinos of two sectors can co-generate both types of matter \cite{16A}, 
naturally leading to $\tilde\Omega_B > \Omega_B$ \cite{16B}.

Dark photons and mirror matter can combine to form a multi-component dark 
matter with properties as diverse as ordinary matter in the universe,
and perhaps astrophysical observations do indicate diverse behavior of dark 
matter in galaxy cluster collisions \cite{32}.

Dark photons and mirror matter help each other to stay hidden from direct 
experimental searches, such as experiments like XENON1T \cite{33}, or in
the search for invisible decays of positronium \cite{34}. Indeed, it is 
clear from (\ref{eq9}) that positronium can oscillate into mirror
positronium through $B_\mu$ or $\tilde B_\mu$ exchange. However, the 
amplitude of this transition is proportional to (assuming ultralight
dark photons, $m_b\ll m_e$)
\begin{equation}
{\cal{A}}(Ps\to\widetilde{Ps})\sim\left (\frac{\epsilon_1}{\sqrt{2}}\,\frac{1}
{\sqrt{1-\epsilon_1^2+\epsilon_2}}\right )^2-\left (\frac{\epsilon_1}
{\sqrt{2}}\,\frac{1}{\sqrt{1-\epsilon_1^2-\epsilon_2}}\right )^2=
-\frac{\epsilon_1^2\epsilon_2}{(1-\epsilon_1^2)^2-\epsilon_2^2}\approx
-\epsilon_1^2\epsilon_2.
\label{eq11}
\end{equation}
Bounds on $\epsilon_1$ from the dark photon searches are quite tight
\cite{35}. It is probably safe to assume that $|\epsilon_1|<10^{-7}$. Then
(\ref{eq11}) is unobservable small for any reasonable value of $\epsilon_2$.

As we are playing with the possibility that at the low energy limit the only
messengers between the ordinary and hidden sectors are their dark photons,
our effective Lagrangian (\ref{eq2}) does not contain photon-mirror photon
mixing. Significant amount of such mixing would imply rather rich 
phenomenology and can provide a straightforward portal for the dark matter 
detection \cite{35A,35B,35C,35D}. However, such mixing is severely limited
experimentally \cite{34}, and from the Big Bang nucleosynthesis, as well as 
from the cosmic microwave background measurements and large scale 
structure considerations \cite{35AA,35BB}.

For the Lagrangian (\ref{eq2}), the photon-mirror photon mixing effectively
is induced due to the dark photon-mirror dark photon mixing. However, this 
effective mixing is naturally small, of the order of $\epsilon_1^2\epsilon_2$,
and constraints on $\epsilon_1$ from the dark photon searches imply that the
mixing is expected to be far below the limits mentioned above.

The situation is different for asymmetric mirror matter scenario, when the 
mirror photon-mirror dark photon mixing parameter $\tilde{\epsilon}_1$ not 
necessarily equals to $\epsilon_1$. Since the mirror photon-mirror dark 
photon and  dark photon-mirror dark photon kinetic mixings are not tightly 
constrained, the induced  photon-mirror photon mixing parameter 
$\epsilon_1\tilde{\epsilon}_1\epsilon_2$ can have a phenomenologically interesting
magnitude of about $10^{-10}$.
 
However, for symmetric mirror matter, it will be impossible to solve the 
small-scale structure puzzles with dissipative mirror dark matter as 
considered in \cite{36}, since such a solution requires 
$\epsilon_1^2\epsilon_2\sim 2\cdot 10^{-10}$ 
\cite{36}\footnote{Perhaps, the small-scale structure puzzle is not a puzzle 
at all for the mirror matter model, although it may be a problem for generic 
dissipative dark matter matter. It was shown in \cite{36AA} that the scale of 
damping of mirror structures depends on the temperature ratio between the 
mirror and ordinary sectors, and for small enough ratio of about 0.1-0.2, 
small scales are suppressed.}. However, this conclusion is correct only if 
there are no stable dark particles charged under the gauge group of dark 
photons, which can play the role of dissipative dark matter in solving 
problems of small-scale structure if $\epsilon_1\sim 10^{-9}$ is not that 
small \cite{36A,36B}. 

Contrariwise, with mirror matter at hand, there is no need to assume 
that dark photons represent a significant fraction of galactic dark matter.
Therefore, dark photon direct-search experiments can only rely on
intense photon sources such as the Sun.

If the dark photon portal is a dominant mechanism connecting visible and mirror 
hidden sectors, and if $\epsilon_1$ and $\epsilon_2$ are both too small, then 
unfortunately experiments to directly search for mirror dark matter will have 
no more chance of success than finding black cat in a dark room.

However, we do not want to become victims of the ``Pygmalion Syndrome'' 
\cite{Synge}, and confuse the real world with what we have imagined here. 
Mirror dark matter search experiments are very important, since only in this
way we can check our models of reality. The pessimistic expectation for the 
results of such experiments is just a testable prediction of the particular 
model discussed in this article.

\section*{Acknowledgments}
We are grateful to Maxim Khlopov, Trevor Searight,  Nirmal Raj, Sunny 
Vagnozzi, and Robert McGehee for useful correspondence. Constructive criticism 
from an anonymous reviewer helped improve the presentation of the manuscript.  
The work is supported by the Ministry of Education and Science of the Russian 
Federation.

\appendix*
\section{Alternative method of diagonalization}
The Referee suggested the following alternative method of diagonalization. 
The method is not simpler than the one given in the main text, but 
nevertheless it is useful for checking calculations, and since its ideology 
is common when considering particle mixing. 

Let's write the Lagrangian (\ref{eq1}) in matrix form: 
\begin{equation}
{\cal L}=-\frac{1}{4}{\bf F}_{\mu\nu}^T{\bf K}{\bf F}^{\mu\nu}+\frac{1}{2}
{\bf A}_\mu^T{\bf M}{\bf A}^\mu-e_a\,{\bf J}_\mu^T{\bf A}^\mu,
\label{eqA1}
\end{equation}
where
\begin{equation}
{\bf K}=\left ( \begin{array}{cccc} 1 & \epsilon_1 & 0 & 0 \\
\epsilon_1 & 1 & 0 & \epsilon_2 \\ 0 & 0 & 1 & \epsilon_1 \\
0 & \epsilon_2 & \epsilon_1 & 1 \end{array} \right ),\;
{\bf M}=\left ( \begin{array}{cccc} 0 & 0 & 0 & 0 \\
0 & m_b^2 & 0 & 0 \\ 0 & 0 & 0 & 0 \\
0 & 0 & 0 & m_b^2 \end{array} \right ),
\label{eqA2}
\end{equation}
and
\begin{equation}
{\bf F}^{\mu\nu}=\left ( \begin{array}{c} F_a^{\mu\nu} \\ F_b^{\mu\nu} \\ 
{\tilde F}_a^{\mu\nu} \\ {\tilde F}_b^{\mu\nu} \end{array} \right ),\;
{\bf A}^\mu=\left ( \begin{array}{c} A_a^\mu \\ A_b^\mu \\ 
{\tilde A}_a^\mu \\ {\tilde A}_b^\mu \end{array} \right ),\;
{\bf J}_\mu=\left ( \begin{array}{c} J_\mu \\ 0 \\ {\tilde J}_\mu \\ 0 
\end{array} \right ).
\label{eqA3}
\end{equation}
The kinetic matrix ${\bf K}$ is reduced to the identity matrix by a linear 
transformation ${\bf A}^\mu={\bf P}{\bf A}^{\prime\mu}$, where
\begin{equation}
{\bf P}=\frac{1}{\sqrt{2}}\left ( \begin{array}{cccc} \frac{\cos{\theta}}
{\sqrt{\lambda_1}} & \frac{\sin{\theta}}{\sqrt{\lambda_2}} & 
\frac{-\cos{\theta}}{\sqrt{\lambda_3}} & \frac{\sin{\theta}}
{\sqrt{\lambda_4}} \\ \frac{-\sin{\theta}}{\sqrt{\lambda_1}} & 
\frac{\cos{\theta}}{\sqrt{\lambda_2}} & \frac{-\sin{\theta}}
{\sqrt{\lambda_3}} & \frac{-\cos{\theta}}{\sqrt{\lambda_4}} \\ 
\frac{\cos{\theta}}{\sqrt{\lambda_1}} & \frac{\sin{\theta}}
{\sqrt{\lambda_2}} & \frac{\cos{\theta}}{\sqrt{\lambda_3}} & 
\frac{-\sin{\theta}}{\sqrt{\lambda_4}} \\ \frac{-\sin{\theta}}
{\sqrt{\lambda_1}} & \frac{\cos{\theta}}{\sqrt{\lambda_2}} & 
\frac{\sin{\theta}}{\sqrt{\lambda_3}} & \frac{\cos{\theta}}
{\sqrt{\lambda_4}} \end{array} \right )={\bf S}\left( \begin{array}{cccc} 
\frac{1}{\sqrt{\lambda_1}} & 0 & 0 & 0 \\
0 & \frac{1}{\sqrt{\lambda_2}} & 0 & 0 \\ 0 & 0 & \frac{1}{\sqrt{\lambda_3}}
& 0 \\  0 & 0 & 0 & \frac{1}{\sqrt{\lambda_4}} \end{array} \right ),
\label{eqA4}
\end{equation}
${\bf S}$ is an orthogonal matrix, 
\begin{eqnarray} &&
\lambda_1=1+\frac{1}{2}\left (\epsilon_2-\sqrt{4\epsilon_1^2+\epsilon_2^2}
\right),\;\;\;\;\;\lambda_2=1+\frac{1}{2}\left (\epsilon_2+\sqrt{4\epsilon_1^2+
\epsilon_2^2}\right), \nonumber \\ &&
\lambda_3=1+\frac{1}{2}\left (-\epsilon_2+\sqrt{4\epsilon_1^2+\epsilon_2^2}
\right),\;\; \lambda_4=1-\frac{1}{2}\left (\epsilon_2+\sqrt{4\epsilon_1^2+
\epsilon_2^2}\right)
\label{eqA5}
\end{eqnarray}
are eigenvalues of the matrix ${\bf K}$, and
\begin{equation}
\tan{2\theta}=\frac{2\epsilon_1}{\epsilon_2},\;\;\;
\cos{2\theta}=\frac{\epsilon_2}{\sqrt{4\epsilon_1^2+\epsilon_2^2}}.
\label{eqA6}
\end{equation}
Note that 
\begin{equation}
{\bf S}= \left ( 
\begin{array}{cccc}  \frac{1}{\sqrt{2}} & 0 & \frac{1}{\sqrt{2}} & 0 \\ 
0 & \frac{1}{\sqrt{2}} & 0 & \frac{1}{\sqrt{2}}  \\ \frac{1}{\sqrt{2}} & 0 & 
\frac{-1}{\sqrt{2}} & 0 \\ 0 & \frac{1}{\sqrt{2}} & 0 & \frac{-1}{\sqrt{2}}  
\end{array} \right ) \left(\begin{array}{rrrr}
\cos{\theta} & \sin{\theta} & 0 & 0 \\
-\sin{\theta} & \cos{\theta} & 0 & 0 \\
0 & 0 & -\cos{\theta} & \sin{\theta} \\
0 & 0 & -\sin{\theta} & -\cos{\theta} \end{array} \right ).  
\label{eqA7}
\end{equation}
The mass matrix ${\bf M}^\prime={\bf P}^T{\bf M}{\bf P}$ ceases to be diagonal 
after the transformation ${\bf P}$:
\begin{equation}
{\bf M}^\prime=m_b^2\left ( \begin{array}{cccc}
\frac{\sin^2{\theta}}{\lambda_1} & -\frac{\sin{\theta}\cos{\theta}}
{\sqrt{\lambda_1\lambda_2}} & 0 & 0 \\ -\frac{\sin{\theta}\cos{\theta}}
{\sqrt{\lambda_1\lambda_2}} & \frac{\cos^2{\theta}}{\lambda_2} & 0 & 0 \\
0 & 0 & \frac{\sin^2{\theta}}{\lambda_3} & \frac{\sin{\theta}\cos{\theta}}
{\sqrt{\lambda_3\lambda_4}} \\ 0 & 0 & \frac{\sin{\theta}\cos{\theta}}
{\sqrt{\lambda_3\lambda_4}} &  \frac{\cos^2{\theta}}{\lambda_4}
\end{array} \right ).
\label{eqA8}
\end{equation}
The upper and lower quarters of this matrix have zero determinants. Therefore,
the eigenvalues of ${\bf M}^\prime$ are $0,\mu^2,0,{\tilde \mu}^2$, where
\begin{eqnarray} &&
\mu^2=m_b^2\left (\frac{\sin^2{\theta}}{\lambda_1}+\frac{\cos^2{\theta}}
{\lambda_2}\right )=\frac{m_b^2}{2\lambda_1\lambda_2}\left [\lambda_1+\lambda_2
+(\lambda_1-\lambda_2)\cos{2\theta}\right], \nonumber \\ &&
{\tilde \mu}^2=m_b^2\left (\frac{\sin^2{\theta}}{\lambda_3}+
\frac{\cos^2{\theta}}{\lambda_4}\right )=\frac{m_b^2}{2\lambda_3\lambda_4}
\left [\lambda_3+\lambda_4+(\lambda_3-\lambda_4)\cos{2\theta}\right].
\label{eqA9}
\end{eqnarray}
Substituting $\lambda$-s from (\ref{eqA5}) and $\cos{2\theta}$ from
(\ref{eqA6}), we recover the relations (\ref{eq8}).

The mass matrix ${\bf M}^\prime$ is diagonalized by an orthogonal 
transformation with the matrix (note that, as follows from  (\ref{eqA5}) and
(\ref{eqA6}), $\lambda_1\cos^2{\theta}+\lambda_2\sin^2{\theta}=1$ and
$\lambda_3\cos^2{\theta}+\lambda_4\sin^2{\theta}=1$)
\begin{equation}
{\bf S_1}=\left ( \begin{array}{rrrr} \sqrt{\lambda_1}\cos{\theta} &
-\sqrt{\lambda_2}\sin{\theta} & 0 & 0 \\ \sqrt{\lambda_2}\sin{\theta} &
 \sqrt{\lambda_1}\cos{\theta}  & 0 & 0 \\ 0 & 0 &  -\sqrt{\lambda_3}
\cos{\theta} & -\sqrt{\lambda_4}\sin{\theta} \\  0 & 0 & \sqrt{\lambda_4}
\sin{\theta} & -\sqrt{\lambda_3}\cos{\theta}\end{array} \right ).
\label{eqA10}
\end{equation}  
The transformation ${\bf S_1}$ does not spoil the canonical form of kinetic
terms, since it is orthogonal.

Photon and mirror photon are degenerate in mass (both massless), therefore,
the corresponding combinations of fields are not uniquely determined (only up 
to an orthogonal transformation between the corresponding fields). In 
(\ref{eq9}) we choose the ``flavor'' basis for them (the photon is connected 
only to the ordinary electric current, and the mirror photon is connected 
only to the mirror electric current). To get the same identification of 
fields, we need one more orthogonal transformation
\begin{equation}
{\bf S_2}=\left ( \begin{array}{cccc} \frac{1}{\sqrt{2}} & 0 & \frac{1}
{\sqrt{2}} & 0 \\ 0 & 1 & 0 & 0 \\  \frac{1}{\sqrt{2}} & 0 & \frac{-1}
{\sqrt{2}} & 0 \\ 0 & 0 & 0 & 1 \end{array} \right ).
\label{eqA11}
\end{equation}
In total, the transformation  ${\bf A}^\mu={\bf P}{\bf S_1}{\bf S_2}{\bf B}^\mu$,
where 
\begin{equation}
{\bf P}{\bf S_1}{\bf S_2}=\left (\begin{array}{cccc} 1 & \frac{-\epsilon_1}
{\sqrt{2(1+\epsilon_2-\epsilon_1^2)}} & 0 & \frac{-\epsilon_1}
{\sqrt{2(1-\epsilon_2-\epsilon_1^2)}} \\ 0 &  \frac{1}{\sqrt{2(1+\epsilon_2-
\epsilon_1^2)}} & 0 & \frac{1}{\sqrt{2(1-\epsilon_2-\epsilon_1^2)}} \\
0 & \frac{-\epsilon_1}{\sqrt{2(1+\epsilon_2-\epsilon_1^2)}} & 1 & 
\frac{\epsilon_1}{\sqrt{2(1-\epsilon_2-\epsilon_1^2)}} \\
0 &  \frac{1}{\sqrt{2(1+\epsilon_2-\epsilon_1^2)}} & 0 & \frac{-1}{\sqrt{2(1-
\epsilon_2-\epsilon_1^2)}} \end{array} \right ), \;\; 
{\bf B}^\mu=\left (\begin{array}{c} A^\mu \\ B^\mu \\ {\tilde A}^\mu \\ 
{\tilde B}^\mu \end{array} \right ), 
\label{eqA12}
\end{equation}
brings the Lagrangian (\ref{eqA1}) exactly in the form described by equations
(\ref{eq7}), (\ref{eq8}) and (\ref{eq9}).

\end{document}